\documentclass[superscriptaddress,twocolumn,showpacs,prb]{revtex4}
\usepackage{graphicx}
\usepackage{tabularx}
\usepackage{amsmath}
\usepackage{amssymb}

\begin{document}

\title{Hysteretic memory effects in disordered magnets} 
                                                       
\author{Helmut G.~Katzgraber}
\affiliation{Theoretische Physik, ETH Z\"urich, 
CH-8093 Z\"urich, Switzerland}

\author{Gergely T.~Zimanyi}
\affiliation{Physics Department, University of California, Davis, 
California 95616, USA}
\date{\today}

\begin{abstract}
We study the return point as well as the complementary point memory effect
numerically with paradigmatic models for random magnets and show that already
simple systems with Ising spin symmetry can reproduce the experimental results
of Pierce {\em et al.}~where both memory effects become more pronounced for
increasing disorder and return point memory is always better than
complementary point memory.
\end{abstract}

\pacs{75.50.Lk, 05.50.+q, 75.40.Mg, 75.60.Nt}
\maketitle

Hysteresis is ubiquitous in nature: it occurs in interacting systems
as a collective phenomenon, in most disordered systems, across first-order 
transitions, in magnetic systems, and in depinning phenomena, among
others. It is also crucial for industrial applications: most notably hysteresis
lies at the very foundation of the magnetic recording industry.
Hysteretic systems are employed as recording media because they retain
their state for a long period after a write operation: they exhibit memory.
Many aspects of this memory effect have been studied: the long-time decay
of information through the superparamagnetic decay~\cite{charap:97} or
how to optimize the signal-to-noise ratio by magnetic modeling, 
among others. 

Because of the central importance of memory in recording media,
other types of memory effects have been studied as well. Pierce {\em et
al}.~have investigated Co/Pt multilayer films by repeated cycling over the
hysteresis 
loop.\cite{pierce:05-ea} 
The macroscopic magnetization returns
to the same value cycle after cycle, a signature of a further type of memory.
It is a profound question whether this memory comes about by the system
returning to the same
{\em microscopic} configuration: an effect dubbed ``return point memory''
(RPM), or whether it is exhibited only on the macroscopic
level. Up to the groundbreaking work of Pierce {\em et al.}, it was not possible 
to address this question experimentally. By using an x-ray magnetic
speckle microscopic technique 
which can indirectly resolve microscopic domain patterns, Pierce {\em et al}.~found the presence of RPM for strong 
disorder, but no RPM for weak disorder. Furthermore, Pierce {\em et al}.~studied 
a second type of memory: whether the system develops the reverse of the
original configuration when the hysteresis sweep reaches the opposite
magnetic field, called ``complementary point memory'' (CPM). They
found that disorder influences CPM in the same way as RPM.
In addition, they report that CPM appears to be smaller than
RPM. However, the RPM--CPM difference does not exceed 10\% and it
is not clear whether the difference could have been caused by
instrumental bias.

The experimental work of Pierce {\em et al}.~spawned two theoretical
studies to date. Deutsch and Mai have
performed micromagnetic simulations using the Landau Lifshitz Gilbert (LLG)
equations~\cite{deutsch:05a} reproducing the above memory
effects. However, only few disorder realizations were considered,
leading to large error bars. Furthermore, the difference between RPM and
CPM was not much bigger than the error bars of the
simulation. Finally, they suggest that the rotation of spins is
primarily responsible for the RPM--CPM difference, i.e., scalar spins
cannot cause the effect (in the absence of random fields). But since
their model Hamiltonian contained numerous terms with several
parameters, further clarification may be useful to determine which of
these terms is the primary cause of the memory effects.
Jagla also used the LLG approach, but for constrained soft
scalar spins, subject to random fields.\cite{jagla:05} He reported
seeing the memory effect and the RPM--CPM difference as well. However, no
quantitative measure of the memory was given and the simulations
were performed only at $T=0$.

Therefore, the following challenges still remain to be understood: 
(i) What is the primary cause of the memory effect, or, equivalently, what is 
the minimal model that exhibits the memory effect? (ii) Does the memory
effect persist at finite temperatures, since the thermal fluctuations
have the potential to destroy microscopic correlations? (iii) What is 
the disorder dependence of the memory effects? (iv) Does the 
RPM--CPM difference convincingly exceed the error bars?
To address these challenges, we have studied RPM and CPM in minimal,
paradigmatic disordered spin models: spin glasses and random-field models. 
Equilibrium and close to equilibrium dynamical properties of spin glasses 
have been studied extensively over the 
years.\cite{binder:86,young:98,kawashima:03} However, 
far-from-equilibrium properties, such as their hysteresis, are much less
understood. Recent studies characterized the hysteresis loop of the
Edwards-Anderson spin glass (SG),\cite{katzgraber:02b-ea}
the random field (RF)
Ising model,\cite{sethna:93} and the Sherrington-Kirkpatrick
model.\cite{pazmandi:99} While the RF model has been studied before, 
a comprehensive study with different models is still lacking.
In this paper we address the above four challenges by studying disordered
Ising-type models with and without frustration.

We first study the nearest-neighbor Edwards-Anderson Ising spin glass
\cite{binder:86} with the Hamiltonian
\begin{equation}
{\mathcal H}_{\rm SG} = 
- \sum_{\langle i, j\rangle} J_{ij} S_i S_j 
- H \sum_i S_i
\label{eq:hameasg}
\end{equation}
in which Ising spins $S_i = \pm 1$ lie on the sites of a square
lattice of size $N = L^2$ with periodic boundary
conditions. The measured quantities show essentially no
size dependence past $L = 20$. The interactions $J_{ij}$ are Gaussian distributed 
with zero mean and standard deviation $\sigma_J$. The simulations are
performed by first saturating the system by applying a large external field $H$
and then reducing $H$ in small steps to reverse the magnetization.
At zero temperature we use standard Glauber dynamics \cite{katzgraber:02b-ea}
where randomly chosen unstable spins (pointing against their local field $h_i =
\sum_{\langle j\rangle} J_{ij} S_j + H$) are flipped until all spins are stable
for each field step. At finite temperatures we perform Monte Carlo simulations 
until the average magnetization is independent of the equilibration
time of the Monte Carlo simulation for each field step. 
All results are averaged over 500 disorder realizations.

\begin{figure}
\begin{center}
\includegraphics[width=0.98\columnwidth]{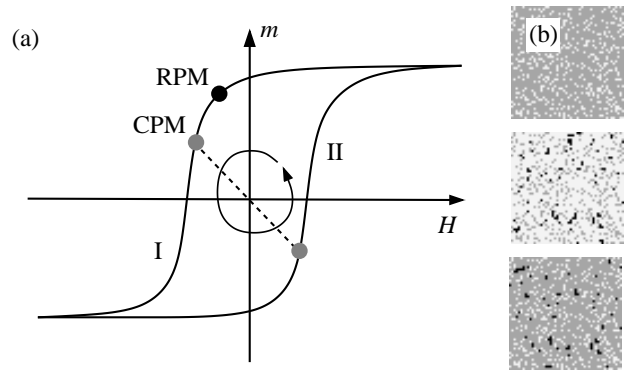}
\end{center}
\vspace*{-0.4cm}
\caption{
(a) In CPM spin configurations on opposite branches I and II 
[$n = 1/2$ in Eq.~(\ref{eq:q})] of the hysteresis
loop (gray dots) are compared, whereas in RPM a configuration on branch I 
[$\{S_i^{(0)}\}$ in Eq.~(\ref{eq:q})] is compared to itself 
after $n \in {\mathbb N}$ cycles around the loop (black dot).
(b) Configurations for $|H^*| = 1.2$ for the two-dimensional 
Edwards-Anderson SG. Light pixels correspond to down spins, dark
pixels correspond to up spins. Top: initial configuration at 
$H^* = -1.2$ (branch I). Center: configuration after a half cycle 
at $H = -H^*$ (CPM, branch II). 
Bottom: configuration at $H = H^*$ after one cycle 
around the hysteresis loop (RPM). The black pixels represent differences 
between the initial and final configurations.}
\label{fig:config}
\end{figure}

Pierce {\em et al}.~captured the RPM and CPM in terms of overlaps of the spin
configurations at different points of the hysteresis loop.
Accordingly, we capture CPM and RPM with $q(H^*)$, the overlap of
the spin configuration $S_i^{(0)}$ at a field $H^*$ with the configuration 
$S_i^{(n)}$ at a field with the same magnitude $|H^*|$ after $n = 1/2$ (CPM) 
and $1$ (RPM) cycles around the hysteresis loop, respectively:
\begin{equation}
q(H^*) = \frac{(-1)^{2n}}{N}\sum_{i = 1}^{N} 
S_i^{(0)}[H^*] S_i^{(n)}[(-1)^{2n}H^*] \; .
\label{eq:q}
\end{equation}
The uniqueness of CPM and RPM is tested by $q'(H,H^*)$, the overlap 
between the spin configuration $S_i^{(0)}$ at $H^*$ with a 
configurations $S_i^{(n)}$ at a field $H$ after $n=1/2$ (CPM) and $1$ (RPM)
cycles around the hysteresis loop: 
\begin{equation}
q'(H,H^*) = \frac{(-1)^{2n}}{N}\sum_{i = 1}^{N} S_i^{(0)}[H^*] S_i^{(n)}[H] \; .
\label{eq:qp}
\end{equation}
To develop a physical picture, Fig.~\ref{fig:config} illustrates 
configurations relevant for the CPM and RPM memory effects. 
The SG displays a large degree of CPM/RPM. 

\begin{figure}
\begin{center}

\includegraphics[width=4.5cm]{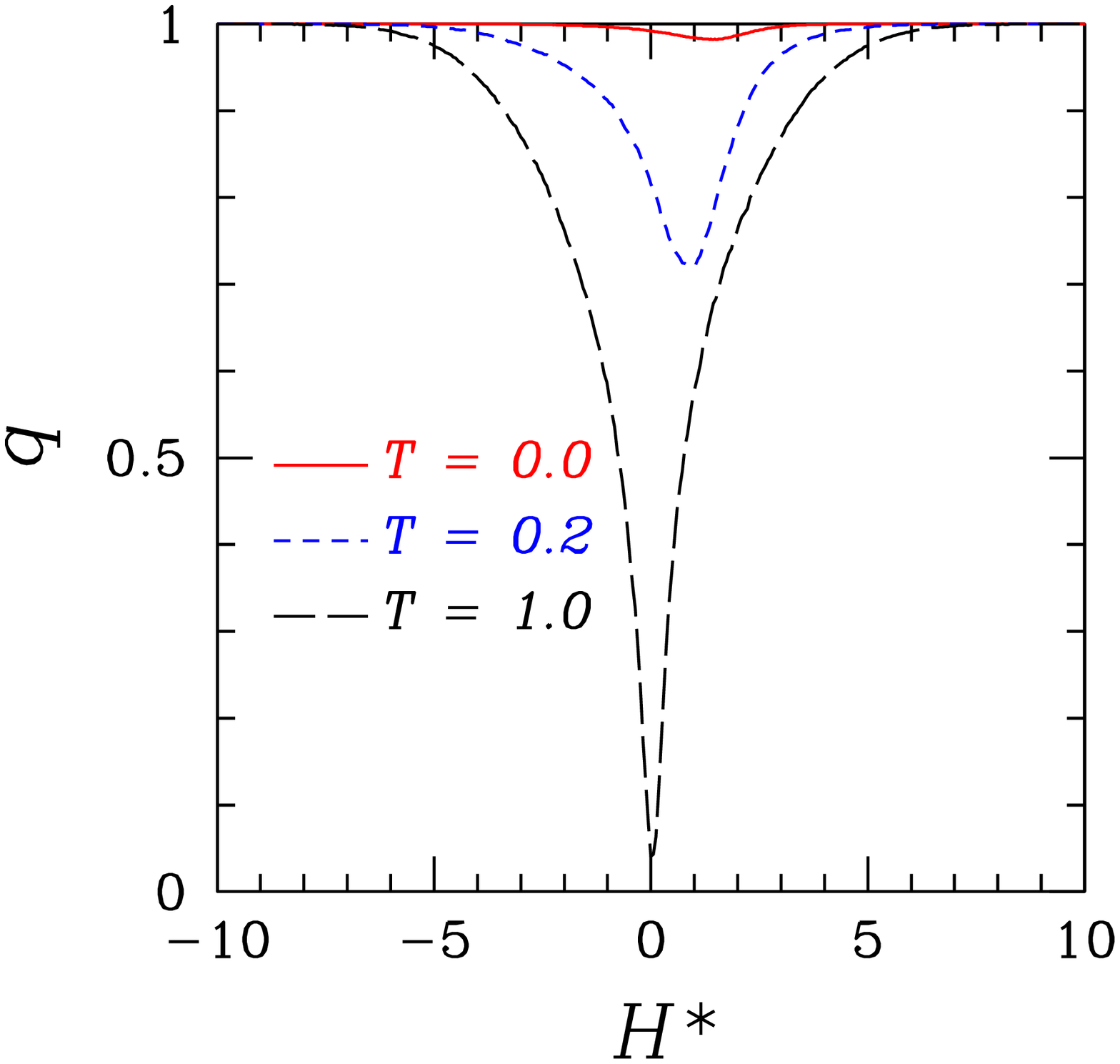}
\hspace*{-0.55cm}
\includegraphics[width=4.5cm]{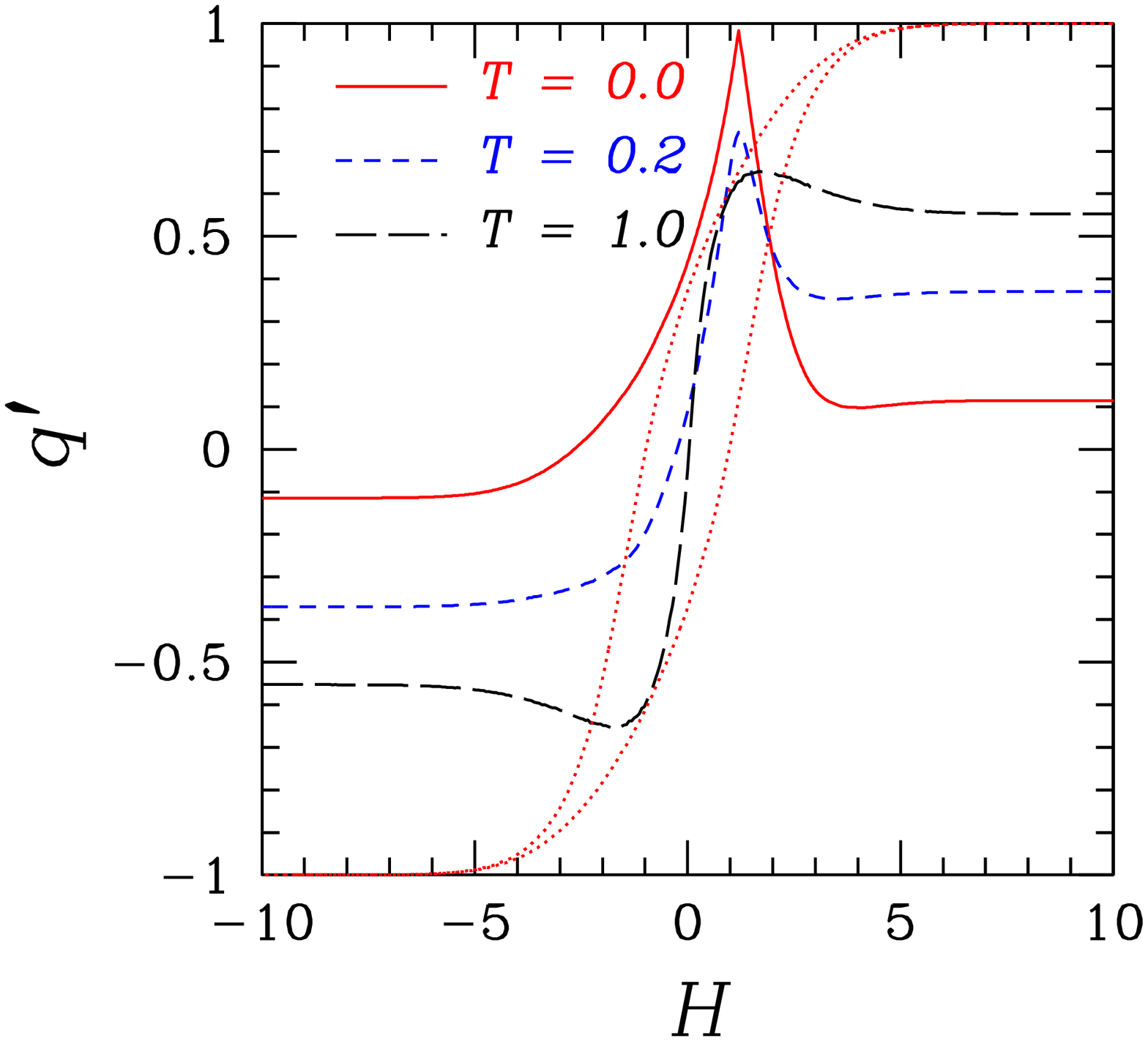}

\end{center}
\vspace*{-0.9cm}
\caption{(Color online)
Overlaps for the 2D Edwards-Anderson SG. 
Left: Overlap $q$ [Eq.~(\ref{eq:q})] for different values of $H^*$ at
different temperatures $T$. Right: Overlap $q'$ [Eq.~(\ref{eq:qp})] 
as a function of the applied field $H$ for $H^* = -1.2$ (coercive field
$H_{\rm c} = -0.98$). The memory becomes
more pronounced for $T \rightarrow 0$. Data for $\sigma_J=1$. The dotted
line represents the zero-temperature major loop.}
\label{fig:easg-qqp}
\end{figure}

Figure \ref{fig:easg-qqp} (left panel) shows $q(H^*)$ 
as a function of $H^*$ for various temperatures. The RPM and CPM curves are 
indistinguishably close. The strong CPM can be attributed to
the spin-reversal symmetry of the system: upon reversing all spins
$S_i$ and the magnetic field $H$, the Hamiltonian transforms into itself. 
Note that RPM is not perfect at $T = 0$ because of the nature of our spin
updating. If unstable spins were sorted and not picked at random, RPM
would be perfect at $T = 0$ (this is not the case if the
ground state of the system is degenerate).
The observation of a robust RPM and CPM memory is an answer to challenge (i), 
establishing the SG as a possible 
minimal model displaying memory effects just as 
in the experiments of Pierce {\em et al}.~with no adjustable parameters. 
Also noteworthy is that the CPM and RPM are smallest around the 
(temperature-dependent) coercive field: as the number of equivalent spin
configurations is the largest in that field region, the
reversal process can evolve along many different paths. 
We also address the temperature dependence of the overlap, as raised 
at (ii). Figure \ref{fig:easg-qqp} shows that the $T=0$ CPM and RPM survive
to finite $T$, even though it would be natural for the thermal fluctuations 
to wash out the memory at the microscopic level 
and convert it to a macroscopic memory only. 
In addition, the memory decreases with increasing temperature.

Figure \ref{fig:easg-qqp} (right panel) shows $q'(H,H^* = -1.2)$ 
for various temperatures. The data show the uniqueness
of CPM and RPM: the overlap function strongly peaks at $H=|H^*|$. Thus, the 
memory is not the result of a gradual slow buildup: there are large-scale 
spin rearrangements during the field sweep in frustrated systems, 
recreating the initial spin configuration only in the close vicinity 
of $H=|H^*|$.
As above, CPM and RPM decrease with 
increasing temperature. The nonzero plateaus are caused by the 
measuring field $H^*=-1.2$ being different from the coercive field, where 
$M(H^*) \neq 0$.  

Next, we address (iii), i.e., we explore the disorder dependence of both
memory effects, since this aspect was an important element of the 
experiments of Pierce {\em et al}. 
\begin{figure}
\begin{center}
                                                                                
\includegraphics[width=4.5cm]{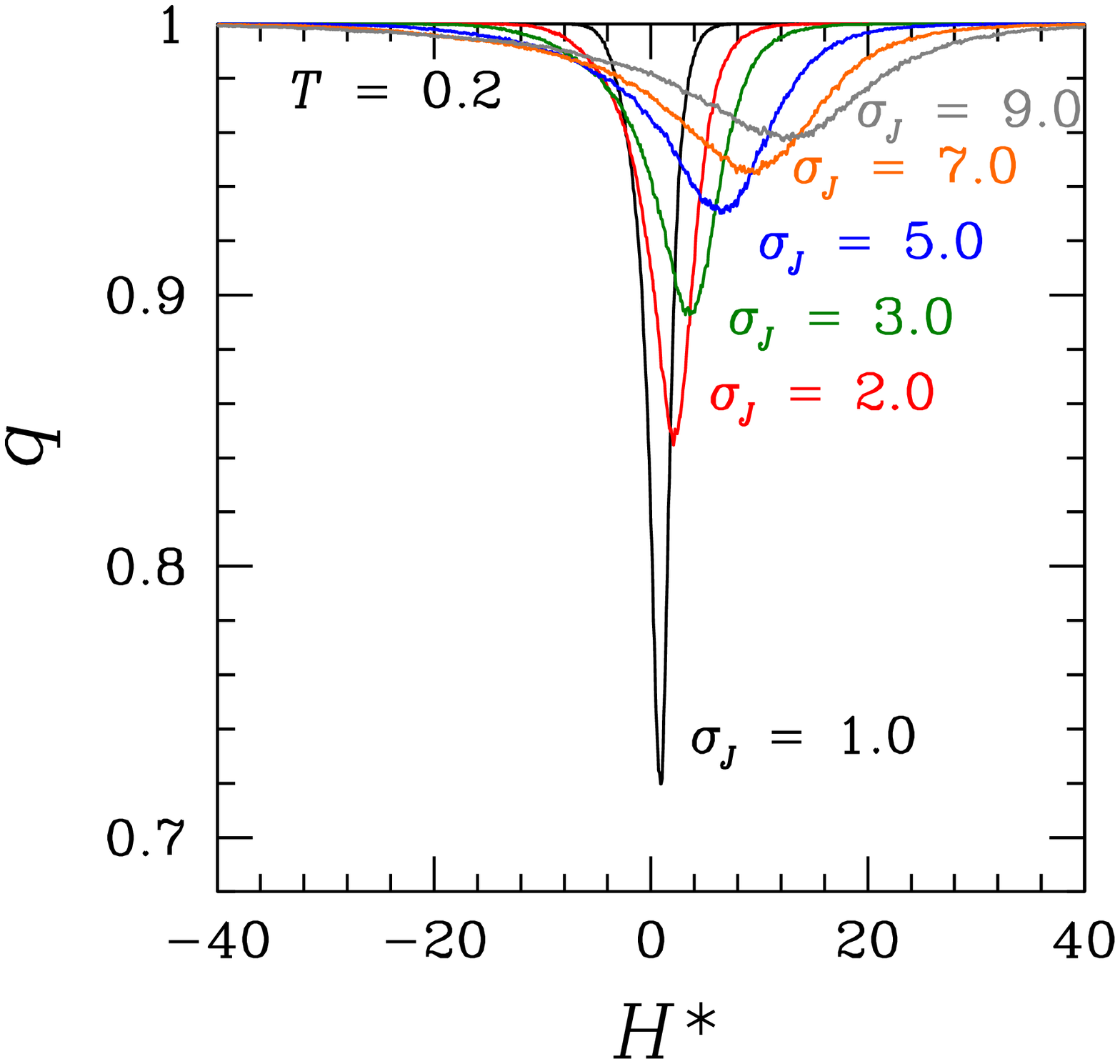}
\hspace*{-0.55cm}
\includegraphics[width=4.5cm]{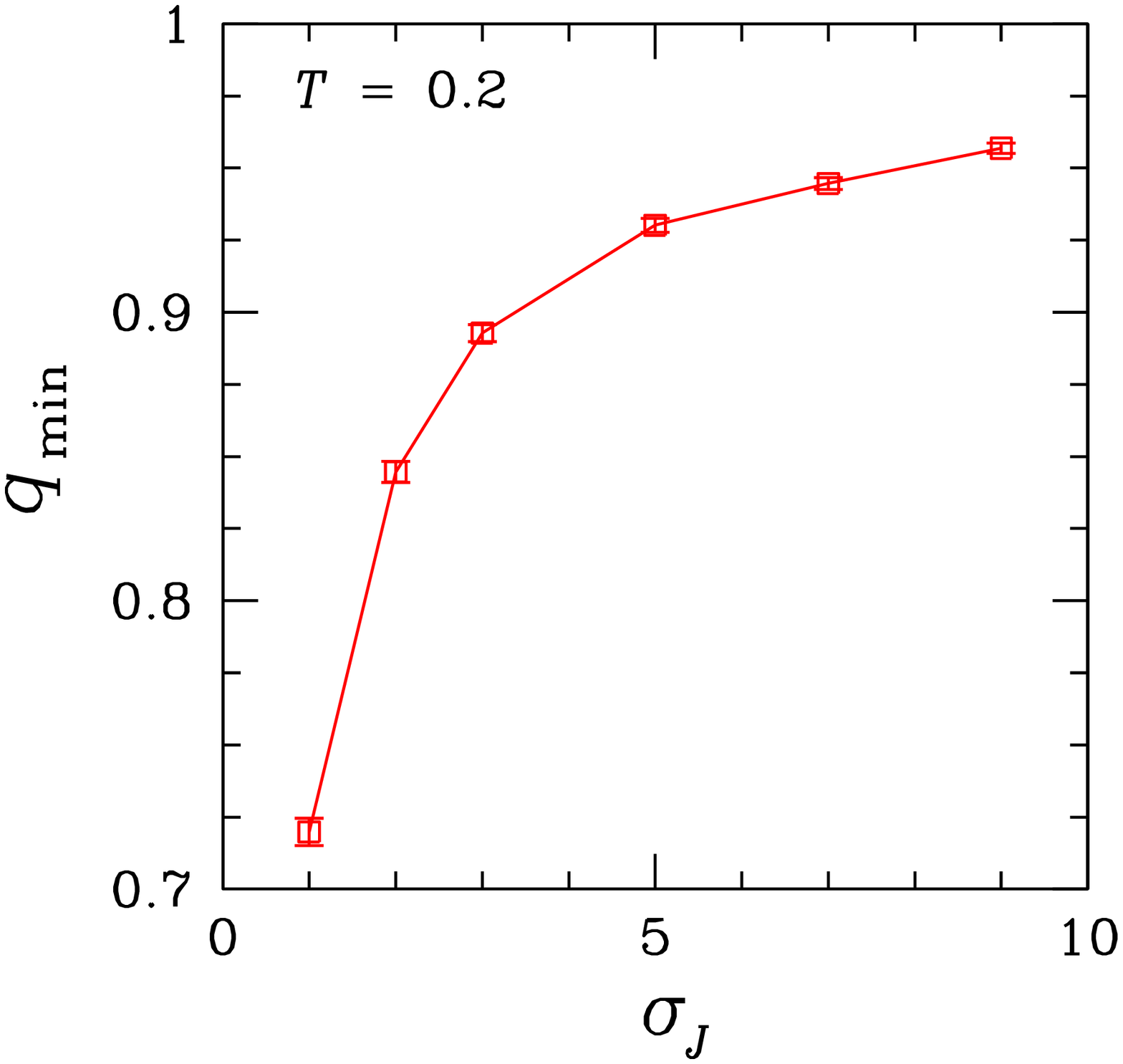}
                                                                                
\end{center}
\vspace*{-0.9cm}
\caption{(Color online)
Left: overlap $q(H^*)$ for RPM and CPM (2D SG) as a function of $H^*$ for
different disorder strengths $\sigma_J$. Right: minimum value 
$q_{\rm min}$ (at coercivity) of $q(H^*)$ as a function of disorder for 
RPM and CPM. 
Both sets of data show that the RPM and CPM increase with increasing 
disorder.}
\label{fig:easg-dis}
\end{figure}
Figure \ref{fig:easg-dis} (left panel) shows $q(H^*)$ as a function
of the disorder $\sigma_J$ at $T=0.2$. The right panel of
Fig.~\ref{fig:easg-dis} shows $q(H^*)$ at coercivity as a function of
$\sigma_J$. RPM and CPM are indistinguishable and visibly
increase with increasing disorder, in good
agreement with Ref.~\onlinecite{pierce:05-ea}. The physical reason
for this correlation is that for weak disorder, the energy landscape 
includes many comparable, shallow valleys, without a single optimal path.
Therefore, during subsequent field sweeps, the system evolves along different 
paths, thus reducing the memory. In contrast, for stronger disorder, 
the energy landscape develops a few preferable valleys and the system evolves
along these optimal valleys during subsequent cycling around the hysteresis 
loop. [The dips in $q(H^*)$ become broader with
increasing disorder because the entire hysteresis loop broadens.] Finally,
in relation to (iv) it is noted that in the SG RPM and CPM are 
indistinguishably close.

Next, we explore RPM and CPM, as well as the same four challenges 
in the 2D random field Ising model,
\begin{equation}
{\mathcal H}_{\rm RF} = - J \sum_{\langle i, j\rangle} S_i S_j 
- \sum_i [H+h_i] S_i \; ,
\label{eq:hamrfim}
\end{equation}
where $J = 1$ is a ferromagnetic coupling and the random fields $h_i$ are
chosen from a Gaussian distribution with zero mean and standard
deviation $\sigma_h$. The main differences between the RF model and the 
Edwards-Anderson SG are that the RF model does not have frustration 
and does not have spin-reversal symmetry.

\begin{figure}
\begin{center}
                                                                                
\includegraphics[width=4.5cm]{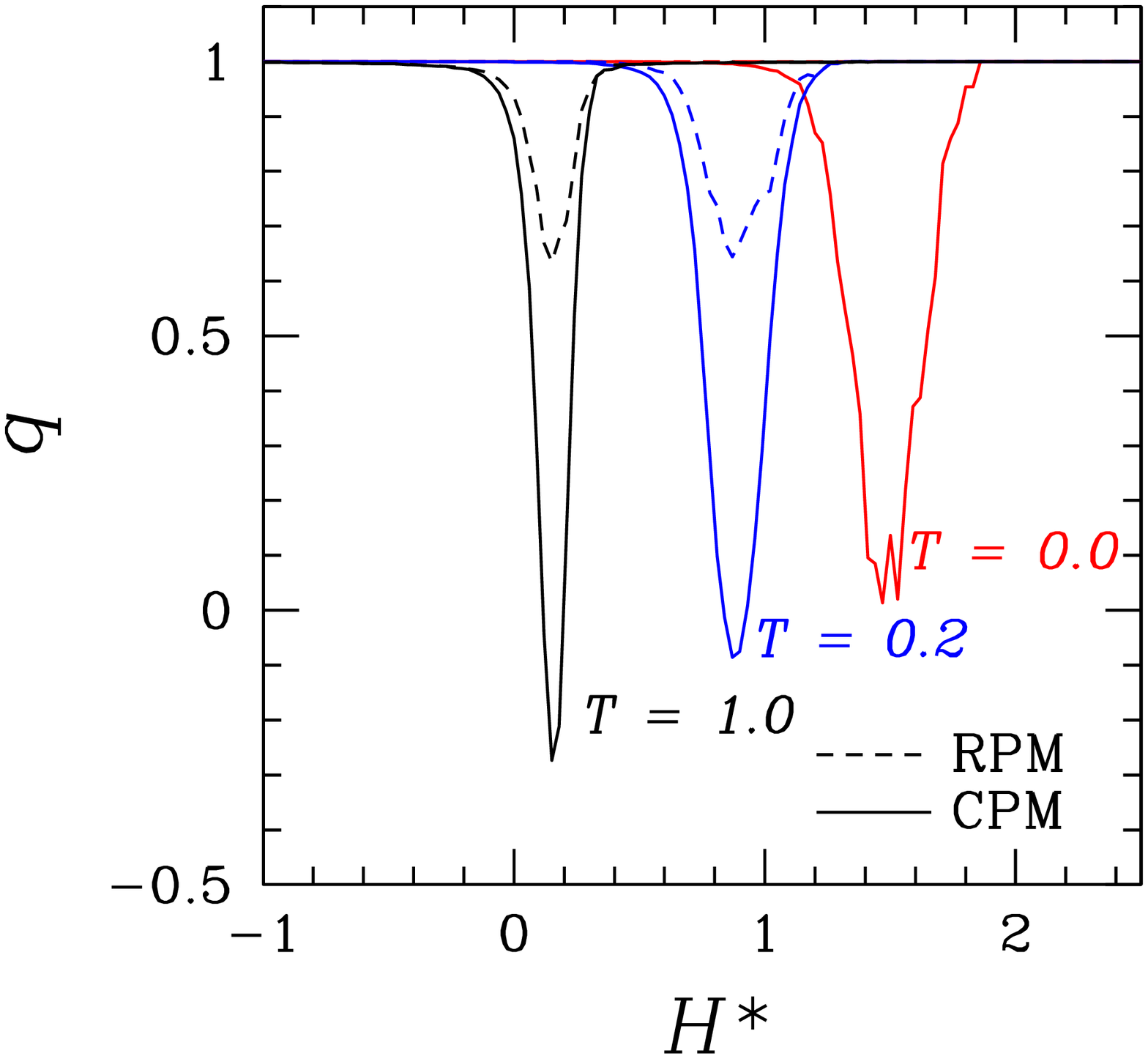}
\hspace*{-0.55cm}
\includegraphics[width=4.5cm]{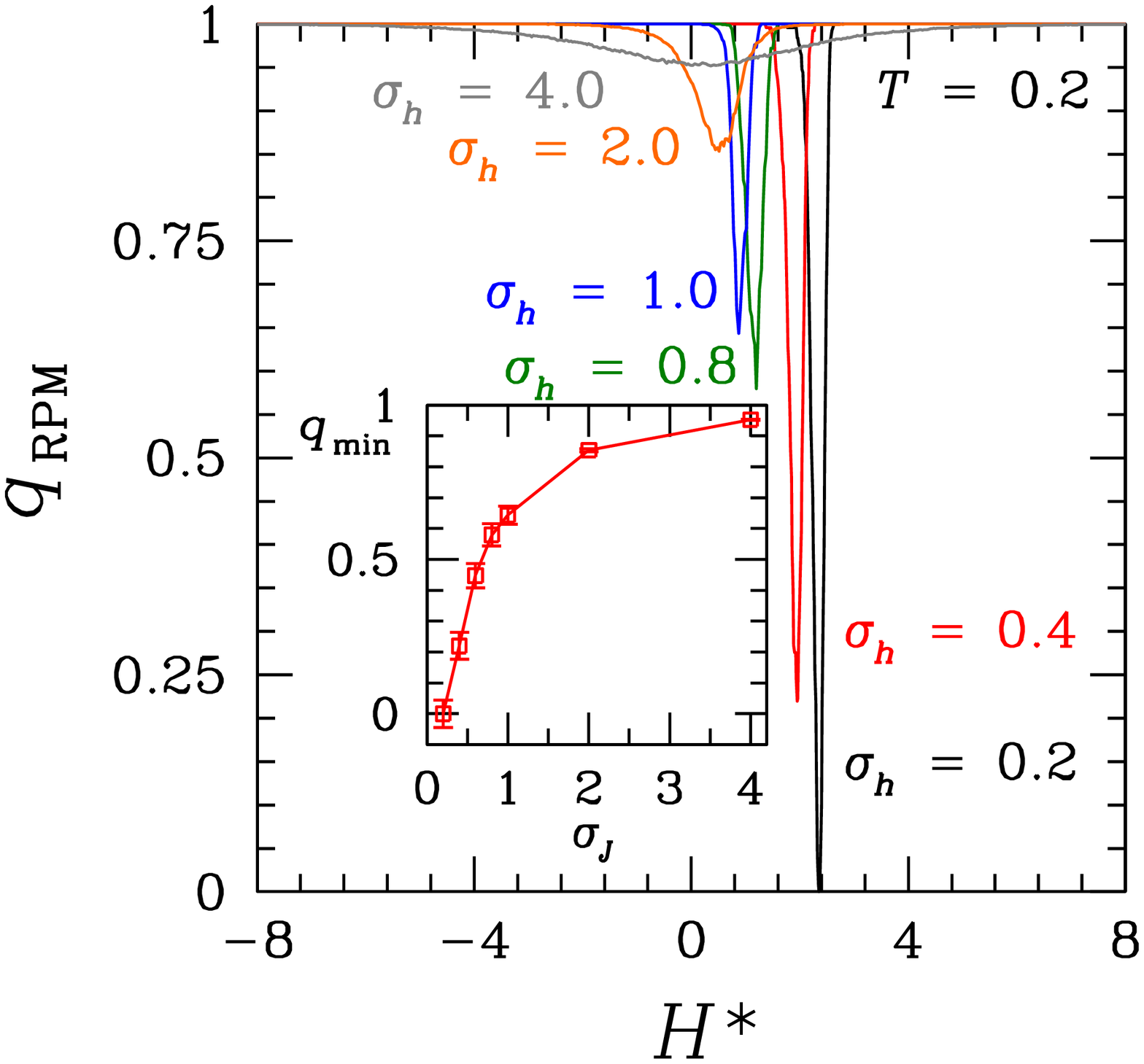}
                                                                                
\end{center}
\vspace*{-0.9cm}
\caption{(Color online)
Left panel: Overlap $q$ for RPM (dashed lines) and CPM
(solid lines) for different temperatures for the RF model.
The RPM is perfect at $T = 0$. For all temperatures the RPM is better than 
the CPM. In particular, for all temperatures the CPM is close to zero at 
coercivity suggesting that CPM is rather poor in this model. Right panel: 
Field dependence of $q(H^*)$ for RPM for different disorder strengths 
$\sigma_h$ ($T = 0.20$). The RPM becomes better with
increasing disorder at finite temperatures. The inset shows the disorder
dependence of $q(H^*)$ for the RPM at coercivity.
}
\label{fig:rfim-dis}
\end{figure}

Figure \ref{fig:rfim-dis} (left panel) shows both $q(H^*)$ for the RPM (dashed
lines) and the CPM (solid lines) as a function of the field $H^*$ for various
temperatures. Concerning (i), this model also exhibits clear 
RPM, whereas CPM is rather poor. Further, (ii) RPM is not washed out by 
thermal fluctuations at once, but is gradually weakened with increasing $T$.

Regarding (iv), the RF model deviates from the SG results and correlates with 
the experiments: in the RF model the RPM and CPM are different. The RPM is 
bigger than the CPM for all temperatures. This effect is due to the fact that 
the RF model does not have spin-reversal symmetry and therefore the spin 
configurations on the ascending branch do not correlate closely with the 
configurations on the descending branch. For intermediate-to-large values of 
the disorder the CPM is negligible and in the proximity of the coercive
field the CPM correlation is in fact negative.
In contrast, the RPM is large in the RF Ising model.  
In particular, at $T=0$ the
RPM is perfect due to the ``no-crossing property'' of the 
RF Ising model.\cite{middleton:92}
Figure \ref{fig:rfim-dis} (right panel) shows (iii) $q(H^*)$ for the RPM for 
different disorder strengths $\sigma_h$ at $T=0.2$.  RPM 
in the RF model also increases with increasing disorder, also illustrated in the
inset: The overlap $q(H^*)$ at coercivity for RPM increases with 
increasing disorder. 
As for the SG, the memory effects increase due to the valleys in the energy 
landscape becoming more pronounced with increasing disorder.
 
In the RF model the CPM is always close to zero, whereas the RPM increases 
with increasing disorder (and is perfect at $T = 0$). Correspondingly, 
the RPM--CPM difference is large over much of the parameter space. 
These findings do not agree completely with the experimental results of 
Pierce {\em et al}. On the other hand the SG lacks the RPM--CPM 
asymmetry completely. Therefore we
have explored whether a combination of the SG and the RF model
might yield results comparable to the experiments: 
increasing memory with increasing disorder, as well as RPM being better 
than CPM. Thus we apply diluted random fields to the SG model:
\begin{equation}
{\mathcal H}_{\rm SG+RF} = - \sum_{\langle i, j\rangle} J_{ij} S_i S_j - 
\sum_i [H + h_i] S_i \; .
\label{eq:hameasgrf}
\end{equation}
The random bonds $J_{ij}$ are chosen from a Gaussian distribution
with zero mean and standard deviation $\sigma_J$. The random fields
are randomly attached to only 5\% of all sites, chosen from a Gaussian 
distribution with zero mean and standard deviation unity. 

\begin{figure}
\begin{center}
                                                                                
\includegraphics[width=4.5cm]{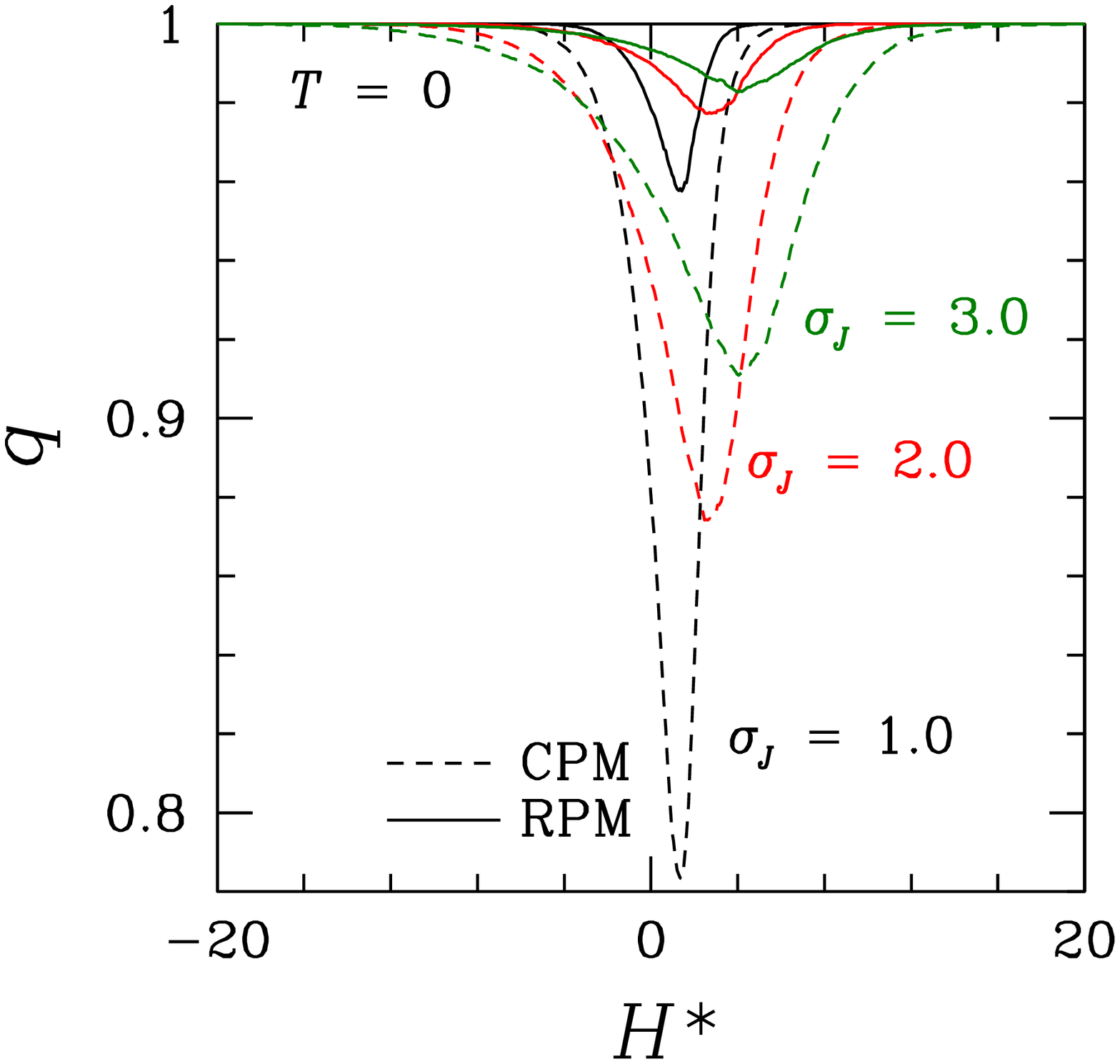}
\hspace*{-0.55cm}
\includegraphics[width=4.5cm]{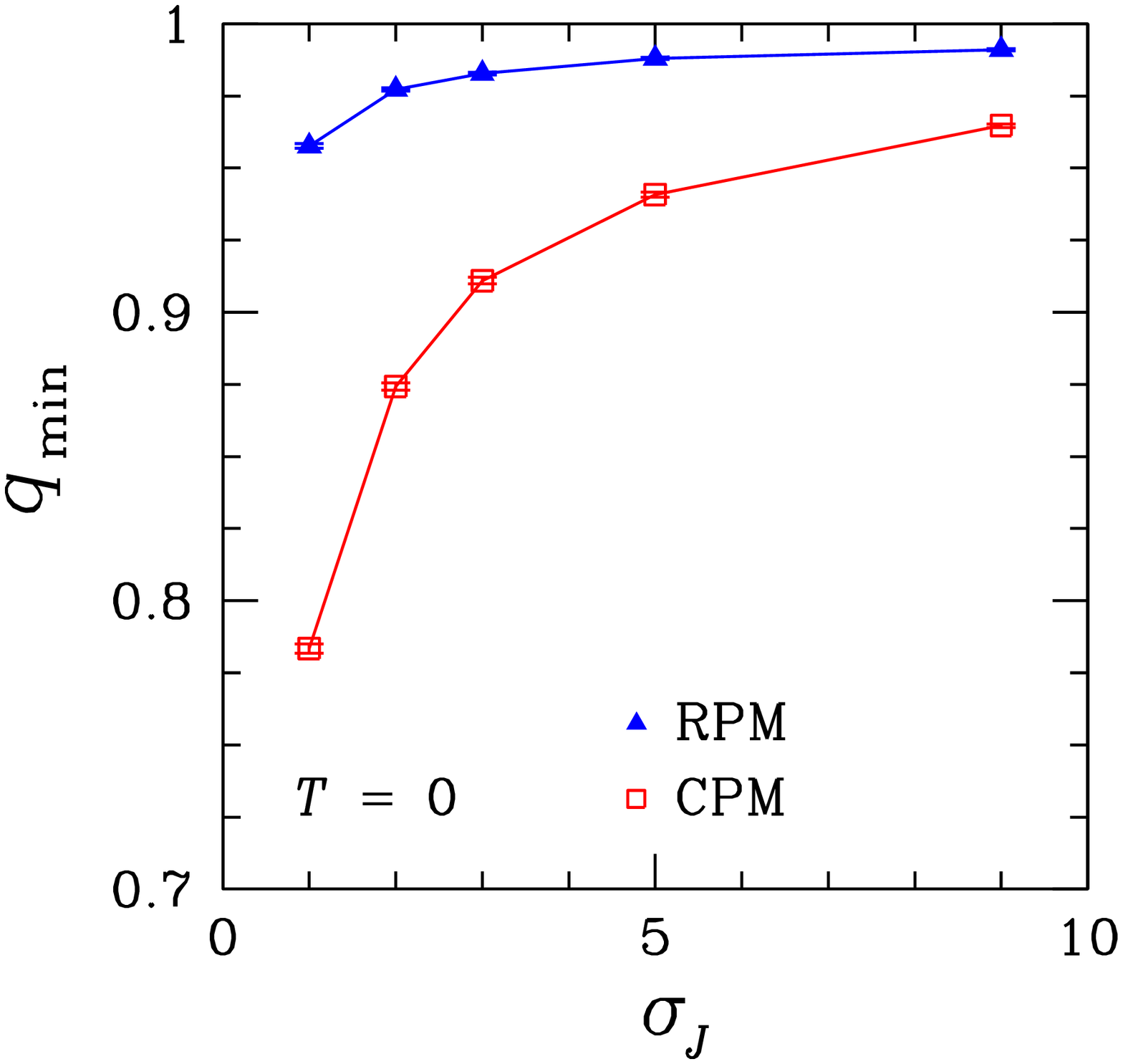}
                                                                                
\end{center}
\vspace*{-0.9cm}
\caption{(Color online)
Left panel: Field dependence of $q(H^*)$ for RPM (solid lines) and
CPM (dashed lines) for different disorder strengths $\sigma_J$ ($T = 0$) for
the SG+RF model [Eq.~(\ref{eq:hameasgrf})]. The RPM is better than the CPM 
and both increase with increasing disorder.
Right panel: Disorder dependence of $q(H^*)$ of the RPM and the CPM at 
coercivity. Both memory effects become better with increasing disorder, 
and RPM is larger than CPM, in good qualitative agreement with the 
experimental results of Pierce {\em et al}.
}
\label{fig:mixed-dis}
\end{figure}
Figure \ref{fig:mixed-dis} illustrates the results at $T=0$ which 
persist well to finite $T$.
The left panel of Fig.~\ref{fig:mixed-dis} shows $q(H^*)$ for both the RPM 
and the CPM as a function of $H^*$ for different disorder strengths $\sigma_J$ 
in the SG+RF model. The right panel of Fig.~\ref{fig:mixed-dis} shows the 
overlap $q$ at coercivity for the RPM and the CPM. Both memory effects 
increase with increasing disorder and the RPM is larger than the CPM, 
in qualitative agreement with experimental results. These findings establish 
that the SG+RF model qualitatively reproduces all aspects of the
Pierce {\em et al} results.

In summary, we have simulated paradigmatic disordered spin models: the 
Edwards-Anderson spin glass, the random-field Ising model, and 
a spin glass with diluted random fields. We have found that (i) all three
models exhibit return point and complementary point 
memory. (ii) Both memory effects persist to finite temperatures.
(iii) Both memory effects increase with increasing disorder. (iv)
In the spin glass the RPM is always identical to the CPM 
because of the spin-reversal symmetry. In the RF Ising model
the CPM is always close to zero because of the lack of spin-reversal 
symmetry. Finally, a spin glass where spin-reversal symmetry is broken with
diluted random fields shows a moderate RPM--CPM 
difference, illustrating the sensitivity of the memory effects to the details
of the models.

In relation to the experiments of Pierce {\em et al}., one 
recalls that the films of Pierce {\em et al}.~have strong out-of plane 
anisotropy, restricting the orientation of the spins to being essentially 
perpendicular to the film. Thus, describing those films in terms of 
Ising Hamiltonians might be a reasonable approximation. 
Furthermore, since the dipolar
interactions in these perpendicular films are antiferromagnetic, they 
introduce extensive frustration into the system, the key ingredient of the 
spin glass model. Finally, spins frozen in by shape anisotropies of a locally 
deformed environment, or by unusually large crystal field anisotropies, or 
possibly by frozen-in reversed bubbles, as reported by Davies {\em et al}.~in 
Co/Pt multilayer films,\cite{davies:04-ea} 
could all be the origin of 
random fields at a few percent of the sites. These considerations make 
it conceivable that all ingredients of the SG+RF model may be present
in the Pierce {\em et al}.~films. In addition, the SG+RF model agrees
qualitatively with the experiments: RPM and CPM increase with increasing
disorder and RPM is always more pronounced than CPM. 

These results, together with the work by Deutsch and Mai, and that of Jagla
show that a RPM-CPM asymmetry can only be obtained when the system's
symmetry is broken. While Deutsch and Mai as well as Jagla break the symmetry
{\em dynamically} using vector models with damped LLG dynamics, in this work we 
present an alternate avenue that breaks the symmetry {\em statically} using 
random fields.
Thus in this work we present a plausible alternative which reproduces all 
aspects of the experiments within a minimal framework.

\paragraph*{Acknowledgments}
We thank J.~M.~Deutsch, E.~E.~Fullerton, O.~Hellwig, Kai Liu,
R.~T.~Scalettar, and L.~Sorensen for useful discussions.

\bibliography{refs}

\end{document}